\documentclass[journal]{IEEEtran}
\usepackage{cite}

\ifCLASSINFOpdf
  \usepackage[pdftex]{graphicx}
\else
\fi
\usepackage{amsmath}
\usepackage{algorithm}
\usepackage{algpseudocode}

\usepackage{array}
\usepackage{stfloats}
\usepackage{color}

\begin{document}
%
\title{Capacity Limitation and Optimization Strategy for Flexible Point-to-Multi-Point Optical Networks}
%
%
%

\author{Ji Zhou,
        Haide Wang,
        Liangchuan Li,
        Weiping Liu,
        Changyuan Yu,
        and Zhaohui Li

\thanks{Manuscript received; revised. This work was supported in part by the  National Natural Science Foundation of China (62371207, 62005102); Hong Kong Scholars Program (XJ2021018). (Corresponding authors: Ji Zhou and Liangchuan Li), (e-mail: zhouji@jnu.edu.cn, liliangchuan@huawei.com).}

\thanks{J. Zhou, H. Wang, and W. Liu are with the Department of Electronic Engineering, College of Information Science and Technology, Jinan University, Guangzhou 510632, China.}

\thanks{Z. Xing and L. Li are with Optical Research Department, Huawei Technologies Co Ltd, Dongguan, 523808, China.}

\thanks{C. Yu is with the Department of Electronic and Information Engineering, The Hong Kong Polytechnic University, Hong Kong.}

\thanks{Z. Li is with the Guangdong Provincial Key Laboratory of Optoelectronic Information Processing Chips and Systems, Sun Yat-sen University, Guangzhou 510275, China}
}

\maketitle

\begin{abstract}
Point-to-multi-point (PtMP) optical networks become the main solutions for network-edge applications such as passive optical networks and radio access networks. Entropy-loading digital subcarrier multiplexing (DSCM) is the core technology to achieve low latency and approach high capacity for flexible PtMP optical networks. However, the high peak-to-average power ratio of the entropy-loading DSCM signal limits the power budget and restricts the capacity, which can be reduced effectively by clipping operation. In this paper, we derive the theoretical capacity limitation of the flexible PtMP optical networks based on the entropy-loading DSCM signal. Meanwhile, an optimal clipping ratio for the clipping operation is acquired to approach the highest capacity limitation. Based on an accurate clipping-noise model under the optimal clipping ratio, we establish a three-dimensional look-up table for bit-error ratio, spectral efficiency, and link loss. Based on the three-dimensional look-up table, an optimization strategy is proposed to acquire optimal spectral efficiencies for achieving a higher capacity of the flexible PtMP optical networks.
\end{abstract}


%
\IEEEpeerreviewmaketitle

\section{Introduction}
\IEEEPARstart{P}{oint}-to-multi-point (PtMP) optical networks with hub-to-leaves architecture become the main solutions for the network-edge applications such as optical access networks and optical metro networks \cite{zhang2023flexible, hosseini2022multi, hosseini2023optimized}. Compared to traditional point-to-point (PtP) optical networks, PtMP optical networks require less optical transceivers, thus have lower capital expense (CapEx) and operating expense (OpEx) costs \cite{back2020capex, campos2023coherent,li2023network}. In current commercial PtMP optical access networks, only time-division multiple access (TDMA) has been widely applied, which is statistical multiple access to make full use of bandwidth \cite{wong2011next, effenberger2019pon, kani2020current}. However, the TDMA-based PtMP optical networks naturally have a high latency, which faces the challenge in the latency-sensitive scenarios \cite{wey2019outlook, kani2017solutions, schaber2023pon}. Particularly, the low-latency requirement becomes stricter and stricter in the future optical networks\cite{dias20235g, larrabeiti2023toward, bidkar2022evaluating}.

Owing to the development of coherent optical techniques, entropy-loading digital subcarrier multiplexing (DSCM) has been commercially applied in the 800Gb/s long-haul optical communications \cite{sun2020800g, welch2022digital, che2018approaching}. Furthermore, coherent entropy-loading DSCM can implement the revolutionary frequency-division multiple access (FDMA) for the PtMP optical networks \cite{welch2021point, xing2023first, 10042001}. For the leaves of the PtMP optical networks, the FDMA can provide the independent frequency channel to achieve low latency, and the entropy loading allocates maximal spectral efficiency to approach high capacity \cite{fan2022point, welch2022digital_1, wang2022record}. Meanwhile, the leaves of FDMA can use a relatively low-bandwidth transceiver matching the subcarrier bandwidth but that of TDMA needs a whole-bandwidth transceiver matching the signal bandwidth. Unfortunately, the high peak-to-average-power-ratio (PAPR) of the entropy-loading DSCM signal causes a low optical power budget, thus limiting the capacity in the peak-power-constrained (PPC) PtMP optical networks \cite{zhou2023clipping, oliveira2023ml, zhou2022100g}.

In this paper, the entropy-loading digital subcarriers with well-matched spectral efficiencies are assigned to the leaves depending on their different link losses, which can approach the highest possible capacity for the flexible PtMP optical networks. Meanwhile, the clipping operation is used to reduce the PAPR of the entropy-loading DSCM signal, thus increasing the optical power budget and gaining more capacity. Furthermore, the theoretical capacity limitation and optimization strategy will be investigated, which can guide the optimal parameter setting for the flexible PtMP optical networks. The main contributions of this paper are as follows:
\begin{itemize}
	\item The theoretical capacity limitation is derived considering the clipping operation for the flexible PtMP optical networks. Meanwhile, an optimal clipping ratio is acquired based on the theoretical capacity limitation.
	\item We establish a three-dimensional look-up table for bit-error ratio (BER), spectral efficiency, and link loss, and propose an optimization strategy to approach the capacity limitation of the flexible PtMP optical networks.
\end{itemize}

The remainder of this paper is organized as follows. The capacity limitation for flexible PtMP optical networks is derived in Section \ref{SectionII}. In Section \ref{SectionIII}, an optimization strategy for flexible PtMP optical networks is introduced in detail. Finally, the paper is concluded in Section \ref{SectionIV}.

\section{Capacity Limitation for Flexible PtMP Optical Networks} \label{SectionII}
\begin{figure}[!t]
\centering
\includegraphics[width=\linewidth]{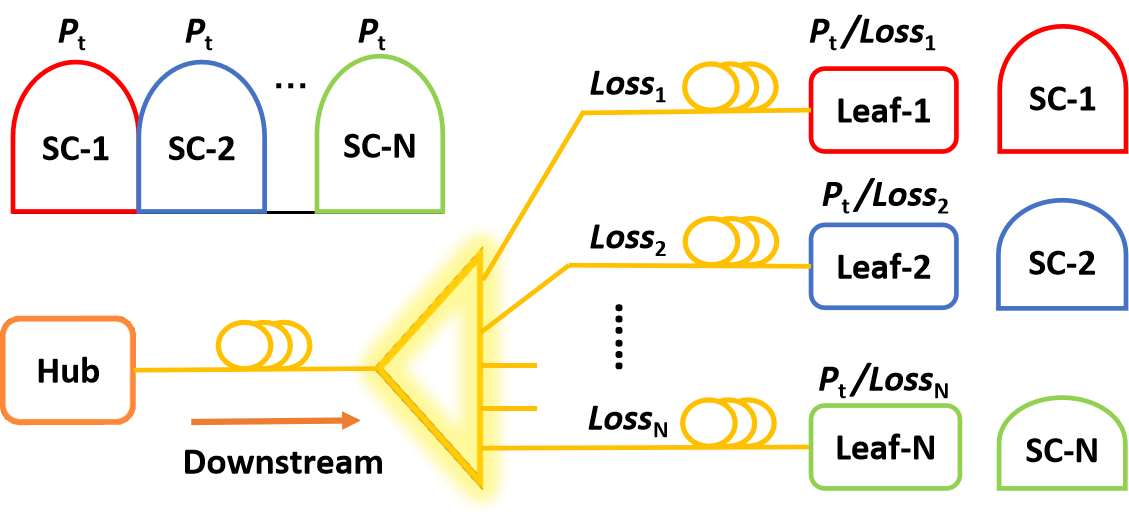}
\caption{The schematic diagram of the DSCM-based flexible PtMP optical networks with different link losses.}
\label{PON-Structure}
\end{figure}

Figure \ref{PON-Structure} shows the schematic diagram of the DSCM-based flexible PtMP optical networks. At the hub, the optical DSCM signal is generated and sent to the leaves. The transmitted subcarriers are set to the same power. After the optical distribution network (ODN), the leaves select and detect their corresponding subcarriers by coherent optical detection. In the actual situation, the optical signals to the different leaves pass the different numbers of optical splitters and different lengths of fiber, leading to different link losses. Based on the on-hand statistic of the deployed PON, the maximum difference of the link losses among the leaves is probably larger than 10 dB \cite{parolari2020flexible}. Therefore, the received signal at the different leaves has different power and signal-to-noise ratio (SNR). By considering the difference of the leaves, we will investigate the theoretical capacity limitation for the DSCM-based flexible PtMP optical networks, and simultaneously use the clipping operation to increase the capacity.

At the hub, the DSCM signal is generated, of which the subcarriers are set to the same average power $P_{\text{t}}$. After the ODN and fiber transmission, the $i$-th subcarrier is selected and detected by the $i$-th leaf. The received signal power of the $i$-th subcarrier can be defined as
\begin{equation} \label{eq1}
P_{\text{r-}i}= P_{\text{t}}/Loss_i = P_{\text{DSCM}}/(N\times Loss_i)
\end{equation}
where ${P_{\text{DSCM}}}$ is the average power of the DSCM signal. $N$ is the subcarrier number in the DSCM signal. $Loss_i$ is the link loss of the $i$-th subcarrier.

Assuming that the devices for the leaves have good uniformity, all the subcarriers suffer from the white noise with the same power. Therefore, the SNR of the DSCM signal on $i$-th subcarrier is expressed as
\begin{equation} \label{eq2}
SNR_i = P_{\text{r-}i} / \sigma^2_\text{n} =  P_{\text{DSCM}}/(N\times Loss_i\times \sigma^2_\text{n})
\end{equation}
where $\sigma^2_\text{n}$ is the variance of the white noise on each subcarrier. Obviously, the increase of ${P_{\text{DSCM}}}$ directly improves the SNR. However, a high-PAPR DSCM signal has a low average power due to the constrained peak power in the PPC PtMP optical networks. The clipping operation can effectively reduce the PAPR of the DSCM signal to increase the average power, which can be expressed as
\begin{equation} \label{eq3}
x_\text{c}=\left\{\begin{array}{cc}
A,&x~>~A\\
x,&\left|x\right|\leq~A \\
-A,&x<-A
\end{array}\right.
\end{equation}
where $A$ is the clipping amplitude. The clipping ratio can be defined as $20\times \text{log}_{10}(\eta)$ where $\eta$ is equal to $A/\sqrt{P_{\text{DSCM}}}$. After the clipping operation, the frequency-domain signal on each subcarrier is expressed as
\begin{equation} \label{eq4}
X_{\text{c}} = \alpha \times X+Noise_{\text{c}}
\end{equation}
where $X$ is the frequency-domain signal on each subcarrier before the clipping operation. $Noise_{\text{c}}$ denotes the clipping noise. The clipping attenuation $\alpha$ can be calculated by 
\begin{equation} \label{eq5}
\alpha = \frac{E(x_\text{c}\cdot x)}{E(x^2)}=\frac{\int_{-\infty}^{\infty} x_\text{c}\cdot x\cdot p(x) \text{d}x}{P_{\text{DSCM}}}= 1-2Q(\eta)
\end{equation}
where $p(x)$ is the probability distribution function of the DSCM signal. Due to the central limit theorem, the DSCM signal with enough subcarriers becomes a Gaussian distribution with zero mean, which can be expressed as
\begin{equation} \label{eq6}
p(x) = \frac{1}{\sqrt{2\pi P_{\text{DSCM}}}}e^{-\frac{x^2}{2P_{\text{DSCM}}}}.
\end{equation}
$Q(x)$ denotes the Q function, which can be defined as
\begin{equation}\label{eq7}
Q(x)=\int_{x}^{\infty}\frac{1}{\sqrt{2\pi}}e^{-\frac{x^2}{2}}dx.
\end{equation}

\begin{figure}[!t]
\centering
\includegraphics[width=0.985\linewidth]{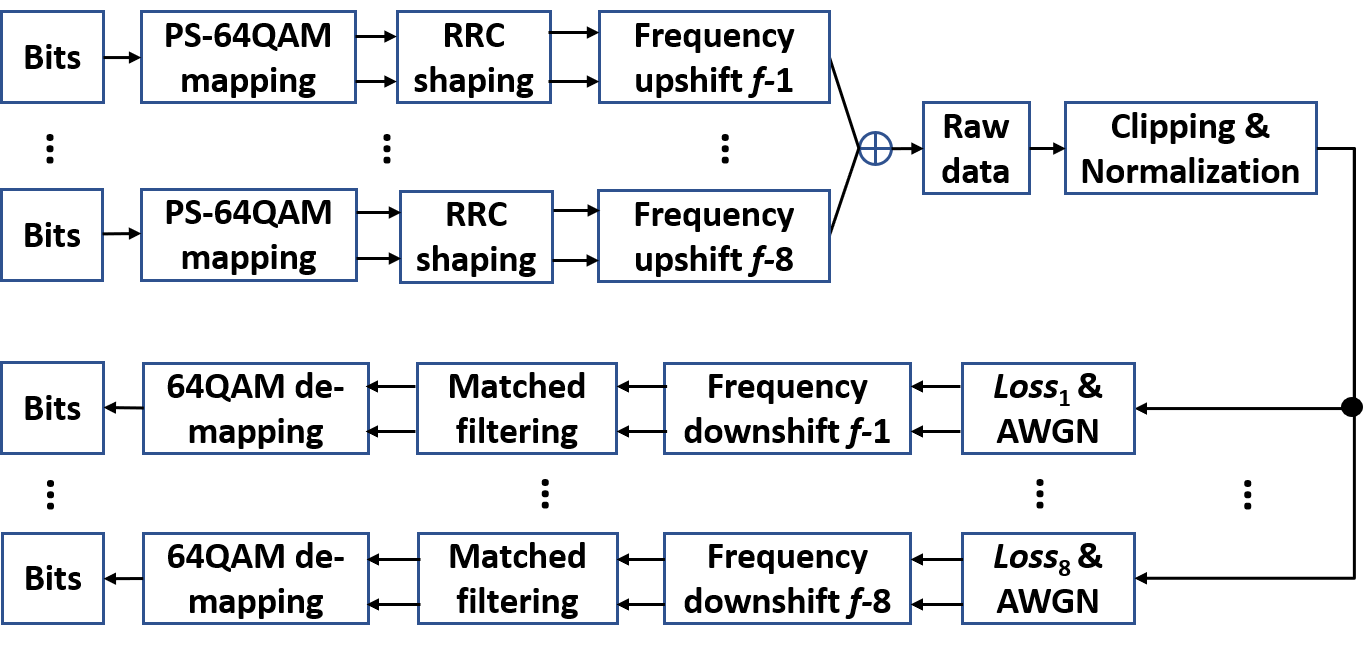}
\caption{The simulation architecture of the flexible PtMP optical networks based on the entropy-loading DSCM.}
\label{SimulationSetup}
\end{figure}

Based on the Eqs. (\ref{eq4}) and (\ref{eq5}), the average power $P_\text{c}$ of the clipping noise $Noise_\text{c}$ can be calculated by
\begin{equation} \label{eq8}
\begin{split}
P_\text{c} &= E(X_\text{c}^2)-\alpha^2E(X^2)\\
&= \int_{-\infty}^{\infty} x_\text{c}^2\cdot p_\text{c}(x)dx-\alpha^2P_{\text{DSCM}}\\
&= 2A^2Q(\eta)+\int_{-A}^{A}x^2p(x)dx-\alpha^2P_{\text{DSCM}}\\
&=2P_{\text{DSCM}}\left\{Q(\eta)[1+\eta^2-2Q(\eta)]-\frac{\eta}{\sqrt{2\pi}}e^{-\frac{\eta^2}{2}}\right\}\\
\end{split}
\end{equation}
where $p_\text{c}(x)$ is the probability distribution of the clipping DSCM signal, which can be defined as
\begin{equation} \label{eq9}
p_\text{c}(x)=\left\{\begin{array}{cc}
p(x), &\left|x\right| \leq A\\
Q(\eta) \delta(\left|x\right|-A), &\left|x\right| > A
\end{array}\right.
\end{equation}
where $\delta(x)$ is the Dirac delta function with a unit impulse.

In the PPC PtMP optical networks, the signal peak is limited to constant value $A_\text{p}$ due to the nonuse of expensive optical amplifier. The peak amplitude $A$ of the clipping DSCM signal is matched to the $A_\text{p}$. We define a matching coefficient $\beta$ as $A_\text{p}/A$ for the clipping DSCM signal. The effective SNR on $i$-th subcarrier of the clipping DSCM signal depends on the power of the clipping signal, clipping noise, and white noise, which can be calculated by
\begin{equation}\label{eq10}
ESNR_{i} = \frac{\alpha^2\times \beta^2\times P_{\text{DSCM}}}{\beta^2\times P_\text{c}+N\times Loss_i\times \sigma_\text{n}^2}.
\end{equation}
Obviously, the $ESNR_i$ depends on the clipping ratio $\eta$ and the link loss $Loss_i$. When the link loss of the $i$-th subcarrier is confirmed, there is an optimal clipping ratio to maximize the effective SNR of the $i$-th subcarrier. The optimal clipping ratios are different among the subcarriers with different link losses. Considering the effective SNRs of all the subcarriers, an optimal clipping ratio can be obtained to maximize the capacity of the DSCM-based flexible P2MP network. The theoretical capacity limitation of the DSCM-based flexible P2MP network can be defined as
\begin{equation} \label{eq11}
C = \sum_{i=1}^{N} C_i =  B\sum_{i=1}^{N} \text{log}_2(1+ESNR_{i})
\end{equation}
where $B$ is the bandwidth of one subcarrier. The optimal clipping ratio to maximize the capacity can be acquired by
\begin{equation} \label{eq12}
\eta_{\text{opt}} = \underset{\eta} {\arg \max}~ B \sum_{i=1}^{N} \text{log}_2(1+ESNR_{i}).
\end{equation}
When the link losses of the leaves are measured, the theoretical results based on Eq. (\ref{eq12}) can get the optimal clipping ratio to achieve the highest capacity for the DSCM-based flexible PtMP optical networks.

\begin{figure}[!t]
\centering
\includegraphics[width=\linewidth]{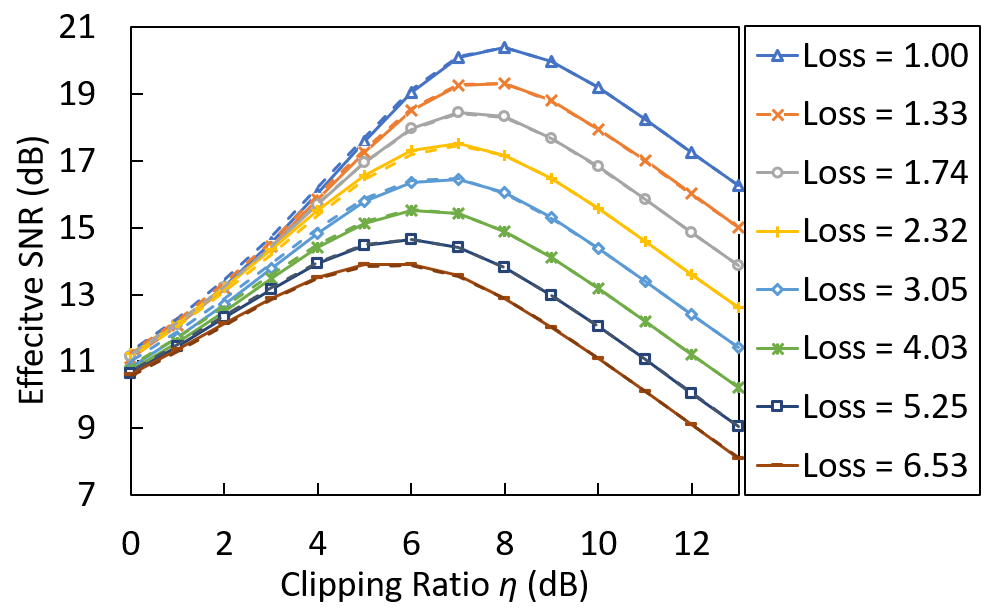}
\caption{The effective SNR versus the clipping ratio for the subcarriers with different link losses. The solid and dashed lines denote the simulation and theoretical results, respectively.}
\label{ESNRvsCR}
\end{figure}

Figure \ref{SimulationSetup} shows the simulation architecture of the flexible PtMP optical networks based on the entropy-loading DSCM. First of all, the SNR of each leaf should be estimated to calculate the spectral efficiency for implementing entropy loading. At the transmitter, the probabilistic-shaping 64-level quadrature amplitude modulation (PS-64QAM) for each subcarrier is mapped depending on the calculated spectral efficiency of each leaf. After the root-raised-cosine-shaping (RRC) shaping and frequency upshift, the PS-64QAM-modulated subcarriers are multiplexed to generate an entropy-loading DSCM signal. The clipping operation is applied in the entropy-loading DSCM signal to reduce the PAPR. After the normalization to the same peak amplitude $A_p$, the different losses and the white noise with the same power are added to the signals on the subcarriers for the different leaves. At the receiver, the frequency downshift and matched filter are used to select the subcarrier and recover the PS-64QAM signal for each leaf. Finally, the recovered PS-64QAM signal is used to estimate the SNR and calculate the BER.

Figure \ref{ESNRvsCR} depicts the effective SNR versus clipping ratio for the subcarriers with different link losses. The solid and dashed lines denote the simulation and theoretical results, respectively. The subcarrier number $N$ is set to 8. The losses of the leaves are set to the $\boldsymbol{Loss}$ of [1, 1.33, 1.74, 2.32, 3.05, 4.03, 5.25, 6.53]. The noise power $\sigma_{\text{n}}^2$ and the peak value $A_p$ are set to 0.0237 and 2.579, respectively. The theoretically effective SNR can be calculated by Eq. (\ref{eq10}). Obviously, the effective SNR obtained by the simulations coincides with the theoretically effective SNR well. The effective SNR is improved with the decrease of clipping ratio at the beginning owing to the increasing signal power and then decreased with the decrease of clipping ratio due to the increasing clipping noise. There is an optimal clipping ratio to maximize the effective SNR for each subcarrier. However, the optimal clipping ratios are different among the subcarriers with different losses.

\begin{figure}[!t]
\centering
\includegraphics[width=0.85\linewidth]{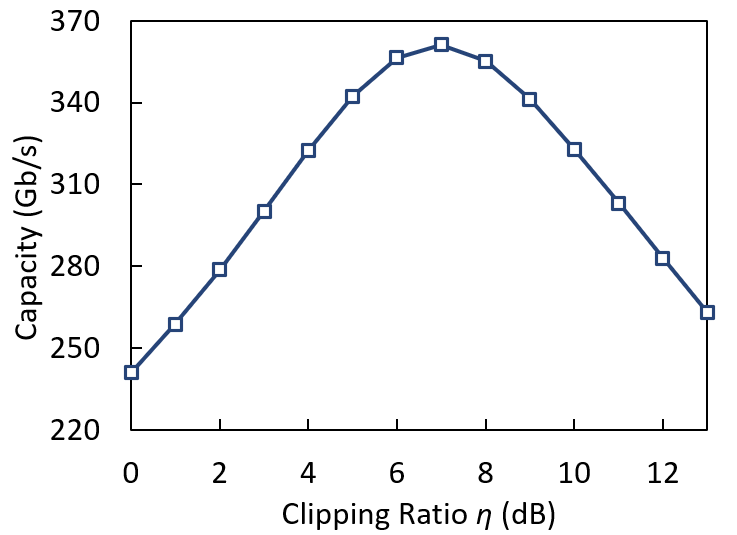}
\caption{The theoretical capacity limitation versus the clipping ratio for the DSCM-based flexible PtMP optical networks when the symbol rate per subcarrier is 8Gbaud.}
\label{CapacityvsCR}
\end{figure}

The theoretical capacity limitation can be calculated by Eq. (\ref{eq11}) for the flexible PtMP network. Meanwhile, the optimal clipping ratio can be obtained by Eq. (\ref{eq12}) to achieve the highest capacity limitation. Fig. \ref{CapacityvsCR} shows the theoretical capacity limitation versus clipping ratio for the DSCM-based flexible PtMP optical networks when the symbol rate per subcarrier is 8Gbaud. When the clipping operation is not used (i.e. a large clipping ratio is set such as 13dB), the capacity limitation of the DSCM-based flexible PtMP optical networks is approximately 263Gb/s. When the clipping operation is used and the clipping ratio is set to the optimal value of $7$dB, the capacity limitation of the DSCM-based flexible PtMP optical networks is approximately 361Gb/s. The capacity limitation of the DSCM-based flexible PtMP optical networks with the optimal clipping operation is approximately 98Gb/s higher than that of the DSCM-based flexible PtMP optical networks without clipping operation.

\section{Optimization Strategy for flexible PtMP Optical Networks} \label{SectionIII}
In this section, we will investigate the optimization strategy to increase the capacity of the DSCM-based flexible PtMP optical networks. A three-dimensional look-up table for BER, spectral efficiency, and link loss will be established to acquire the spectral efficiencies for the subcarriers with different losses under the same targeted BER.

\begin{figure}[!t]
\centering
\includegraphics[width=\linewidth]{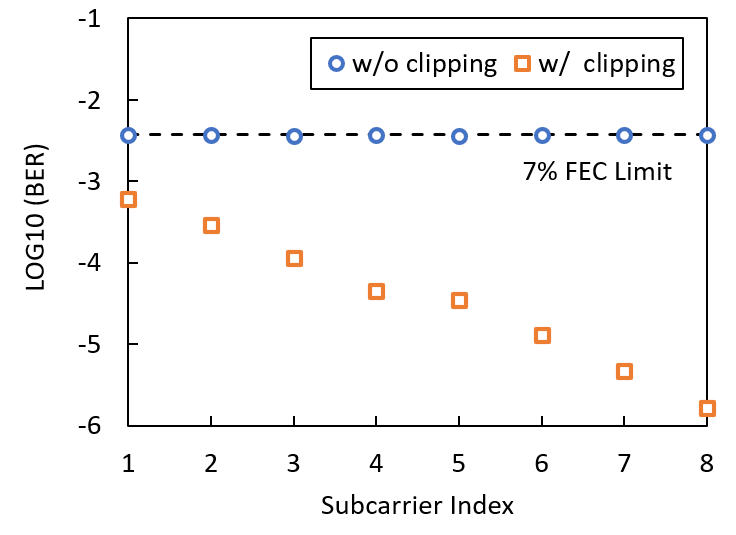}
\caption{The BERs of the subcarriers for the DSCM-based flexible PtMP optical networks without and with the clipping operation.}
\label{BERperSubcarrier}
\end{figure}

Figure \ref{BERperSubcarrier} depicts the BERs of the subcarriers for the DSCM-based flexible PtMP optical networks without and with the clipping operation. When the clipping operation is not employed, the spectral efficiencies of the 8 subcarriers are set to the $\boldsymbol{SE}$ of [4.80, 4.40, 4.00, 3.60, 3.20, 2.80, 2.40, 2.00] bits/symbol depending on the $\boldsymbol{Loss}$ of [1, 1.33, 1.74, 2.32, 3.05, 4.03, 5.25, 6.53]. Under the $\boldsymbol{SE}$, the BER of each subcarrier is on the forward error correction (FEC) limit with 7\% overhead. When the clipping operation is used and the clipping ratio is set to optimal 7dB, the BER performance of each subcarrier is much improved, particularly the subcarrier with low spectral efficiency. To obtain a higher capacity, the spectral efficiency of each subcarrier can be increased until the BER approaches the 7\% FEC limit. In this section, we will establish a three-dimensional look-up table for the BER, spectral efficiency, and link loss. When the link losses of the leaves and targeted BERs of the subcarriers are confirmed, the corresponding spectral efficiency for each subcarrier can be obtained from the three-dimensional look-up table to achieve the highest capacity for the DSCM-based flexible PtMP optical networks.

\begin{figure}[!t]
\centering
\includegraphics[width=\linewidth]{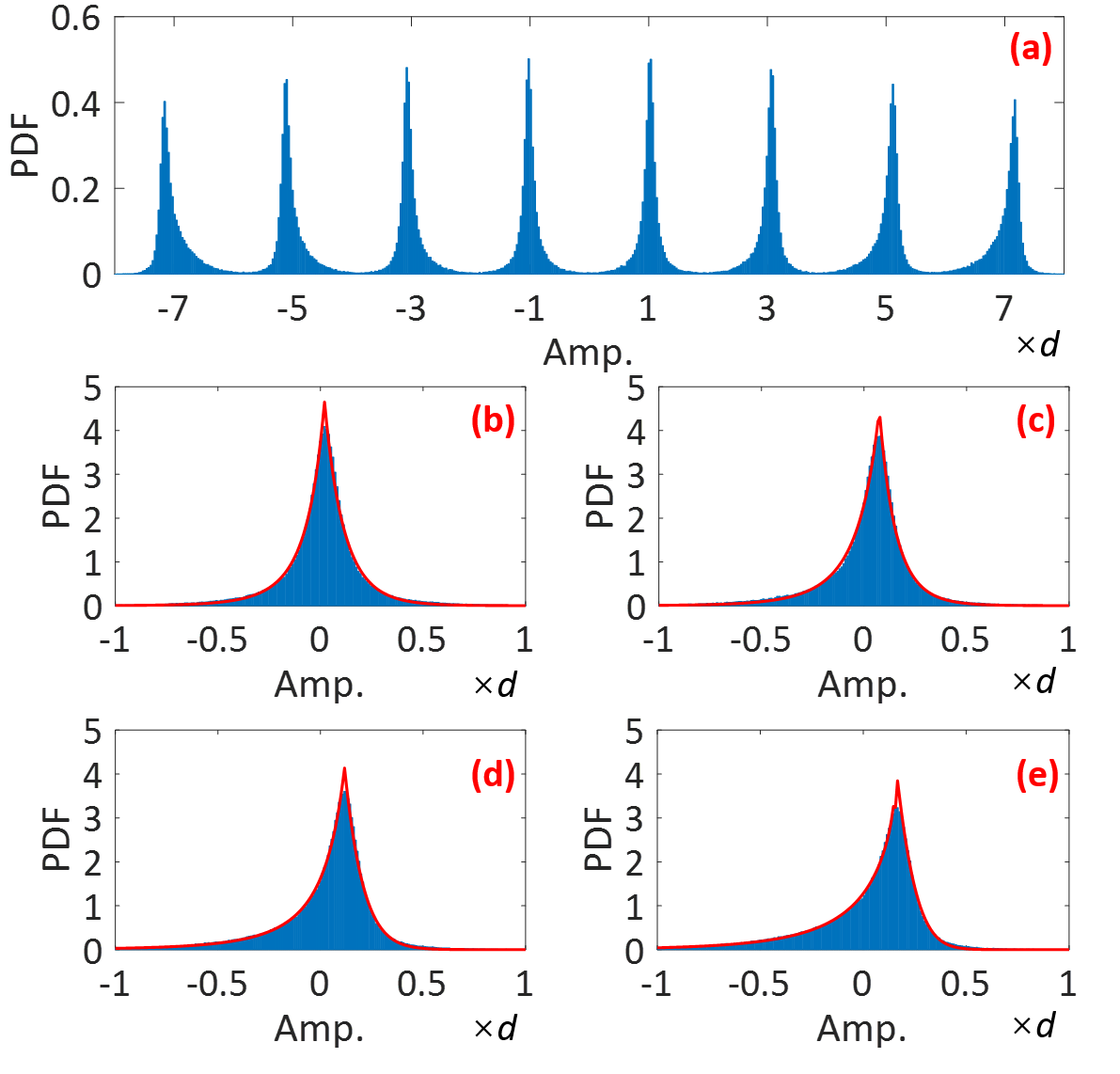}
\caption{(a) Probability density function (PDF) of the in-phase PS-8PAM of the PS-64QAM in the clipping entropy-loading DSCM signal. The Probability density function of clipping noise on the amplitude of (b) 1, (c) 3, (d) 5, and (e) 7, respectively. The red lines denote the fitting curves based on the piecewise power exponential model.}
\label{PDFPAM8}
\end{figure}

For obtaining the three-dimensional look-up table, a theoretical BER should be derived considering the white noise and clipping noise. The PS-64QAM signal can be decomposed into independent in-phase and quadrature amplitudes, which are PS 8-level pulse amplitude modulation (PS-8PAM) signals. Gray code is applied to generate the PS-8PAM signals. Fig. \ref{PDFPAM8} (a) shows the probability density function for the in-phase PS-8PAM of the PS-64QAM in the clipping entropy-loading DSCM signal. It demonstrates that the clipping noise does not coincide with the Gaussian distribution. Figs. \ref{PDFPAM8} (b)-(e) depict the probability density functions of clipping noise on the amplitudes of 1, 3, 5, and 7, respectively. Obviously, the distributions of the clipping noise on different amplitudes are different. The piecewise power exponential model is used to get the accurate fitting curves for the probability density function of the clipping noise on the different amplitudes as the red lines in Figs. \ref{PDFPAM8} (b)-(e) show. Based on the piecewise power exponential model, the theoretical probability density function of the clipping noise can be defined as
\begin{equation} \label{eq13}
f_{Y_\text{c}}(y \mid k)=\left\{\begin{array}{cc}
A_{k,1}\times e^{\frac{-\left|y-\mu_{k,1}\right|^{b_{k,1}}}{2\sigma_{k,1}^2}},~&y \leq D_{k} \\A_{k,2}\times e^{\frac{-\left|y-\mu_{k,2}\right|^{b_{k,2}}}{2 \sigma_{k,2}^2}},~&y>D_{k}
\end{array}\right.
\end{equation}
where $k$ is the amplitudes of [$-7$, $-5$, $-3$, $-1$,~1,~3,~5,~7] for the in-phase amplitude of the PS-64QAM in the clipping entropy-loading DSCM signal. $D_{k}$ is the value corresponding to the highest probability near the amplitude of $k$. $A_{k,1}$, $\mu_{k,1}$, $b_{k,1}$ and $\sigma_{k,1}^2$ are the fitting parameters of the signal at the left of $D_{k}$, while $A_{k,2}$, $\mu_{k,2}$, $b_{k,2}$ and $\sigma_{k,2}^2$ are the those of the signal at the right of $D_{k}$, respectively. 

Based on Eq. (\ref{eq13}), the accurate fitting curves are used to acquire the fitting parameters of $A_{k,1}$, $\mu_{k,1}$, $b_{k,1}$, $\sigma_{k,1}^2$, $A_{k,2}$, $\mu_{k,2}$, $b_{k,2}$, $\sigma_{k,2}^2$, and $D_{k}$ for the theoretical probability density function of the clipping noise. Meanwhile, to make sure the integral value of $f_{Y_{\text{c}}}(y\mid k)$ is equal to $1$, the $A_{k,i}$ should be updated by 
\begin{equation} \label{eq14}
A_{k, i}=\frac{A_{k, i}}{\int_{-\infty}^{\infty} f_{Y_\text{c}}(y \mid k) \text{d} y}
\end{equation}
where $i = 1, 2$ denotes the left and the right parts of the probability density function, respectively.

The theoretical BER can be derived by the joint probability density function based on the probability density functions of the white noise and clipping noise. The white noise and clipping noise can be considered as two independent continuous random variables $Y_\text{n}$ and $Y_\text{c}$. The white noise coincides with Gaussian distribution, and its probability density functions $f_{Y_{\text{n}}}$ is equal to $1/(\sqrt{2\pi}\sigma_\text{n})e^{-y^2/(2\sigma_\text{n}^2)}$. The probability density functions of the clipping noise $f_{Y_{\text{c}}}$ is equal to Eq. (\ref{eq13}).  The random variable of combined noise $Z$ is equal to $Y_\text{n}+Y_\text{c}$, and its probability density functions $f_Z$ can be calculated by
\begin{equation}\label{eq15}
f_Z(z) = \int_{-\infty}^{\infty}f_{Y_{\text{n}}}(y)f_{Y_{\text{c}}}(z-y) \text{d} y
\end{equation}
Substituting $f_{Y_{\text{n}}}$ and $f_{Y_{\text{c}}}$ into Eq. (\ref{eq15}), the probability density functions $f_Z$ can be expressed as
\begin{equation}\label{eq16}
\begin{aligned} 
f_Z(z \mid k) &=\frac{A_{k,1}}{\sqrt{2 \pi} \sigma_\text{n}} \int_{z-D_{k}}^{+\infty} e^{-\frac{y^2}{2 \sigma_\text{n}^2}}e^{-\frac{\left|z-y-\mu_{k,1}\right|^{b_{k,1}}}{2 \sigma_{k,1}^2}} \text{d} y \\
&+\frac{A_{k,2}}{\sqrt{2 \pi} \sigma_\text{n}} \int_{-\infty}^{z-D_{k}} e^{-\frac{y^2}{2 \sigma_\text{n}^2}}e^{-\frac{\left|z-y-\mu_{k,2}\right|^{b_{k,2}}}{2 \sigma_{k,2}^2}} \text{d} y
\end{aligned}
\end{equation}

The BER of the PS-64QAM signal is similar to that of the decomposed PS-8PAM signal. For the PS-8PAM signal, the probabilities of the amplitudes are different, which can be defined as $\boldsymbol{Pr}=[Pr_{\pm1}, Pr_{\pm3}, Pr_{\pm5}, Pr_{\pm7}]$. The theoretical error ratio of the first bit for the PS-PAM8 signal can be calculated by
\setcounter{equation}{16}
\begin{equation}
\begin{aligned}
E_{\text{b,~1}} & = 2\left[Pr_1\int_{-\infty}^{-d} f_Z(z\mid 1) \text{d} z+Pr_3 \int_{-\infty}^{-3d} f_Z(z\mid 3) \text{d} z\right. \\
& \left.+Pr_5 \int_{-\infty}^{-5d} f_Z(z \mid 5) \text{d} z+Pr_7 \int_{-\infty}^{-7 d} f_Z(z \mid 7) \text{d} z\right]
\end{aligned}
\label{eq17}
\end{equation}
where $d$ is the Euclidean distance. Similarly, the theoretical error ratio of the second bit can be calculated by
\begin{equation}
\begin{aligned}
E_{\text{b},~2} & = 2Pr_1 \left[\int_{3 d}^{\infty} f_Z(z\mid 1) \text{d} z+\int_{-\infty}^{-5 d} f_Z(z\mid 1) \text{d} z\right] \\
& +2Pr_3 \left[\int_{d}^{\infty} f_Z(z\mid 3) \text{d} z+\int_{-\infty}^{-7 d} f_Z(z\mid 3) \text{d} z\right] \\
& +2Pr_5 \int_{-9 d}^{ -d} f_Z(z\mid 5) \text{d} z+2Pr_7 \int_{-11 d}^{3 d} f_Z(z\mid 7) \text{d} z.
\end{aligned}
\label{eq18}
\end{equation}
The theoretical error ratio of the third bit can be calculated by Eq. (\ref{eq19}). 
\newcounter{TempEqCnt1}
\setcounter{TempEqCnt1}{\value{equation}}
\setcounter{equation}{18}
\begin{figure*}[!t]
\begin{equation} \label{eq19}
\begin{aligned}
E_{\text{b},~3} & = 2 \left\{Pr_1 \left[\int_{d}^{5 d} f_Z(z\mid 1) \text{d} z+\int_{-7 d}^{-3 d} f_Z(z\mid 1) \text{d} z\right]  +Pr_3 \left[\int_{3d}^{\infty} f_Z(z \mid 3) \text{d} z+\int_{-5 d}^{-d} f_Z(z\mid 3) \text{d} z+\int_{-\infty}^{-9 d} f_Z(z \mid 3) \text{d} z\right] \right.\\
& \left.+Pr_5 \left[\int_{d}^{\infty} f_Z(z \mid 5) \text{d} z+\int_{-7 d}^{-3d} f_Z(z \mid 5) \text{d} z+\int_{-\infty}^{-11 d} f_Z(z \mid 5) \text{d} z\right] +Pr_7 \left[\int_{-5d}^{-d} f_Z(z\mid 7) \text{d} z+\int_{-13 d}^{-9 d} f_Z(z \mid 7) \text{d} z\right] \right\}
\end{aligned}
\end{equation}
\hrulefill
\end{figure*}
\setcounter{equation}{\value{TempEqCnt1}}

As shown in Eqs. (\ref{eq17})-(\ref{eq19}), the $d$ should be confirmed before calculating the theoretical error ratios of the three bits for the PS-8PAM signal. The relationship between $d$ and the average power of the PS-64QAM signal can be defined as 
\setcounter{equation}{19}
\begin{equation}\label{eq20}
d = \sqrt{\frac{P_\text{PS-64QAM}}{4\sum_{i=0}^3 (2i+1)^2\times Pr_{2i+1}}}.
\end{equation}
where $P_\text{PS-64QAM}$ is the average power of the PS-64QAM signal on the subcarrier, which can be calculated by $\alpha^2\times\beta^2\times P_{\text{DSCM}}/(N\times Loss_i)$. Finally, the theoretical BER of the PS-64QAM signal on the subcarrier can be calculated by
\begin{equation} \label{eq21}
E_{\text{b}}=(E_{\text{b, 1}}+E_{\text{b, 2}}+E_{\text{b, 3}})/3.
\end{equation}

\begin{figure}[!t]
\centering
\includegraphics[width=\linewidth]{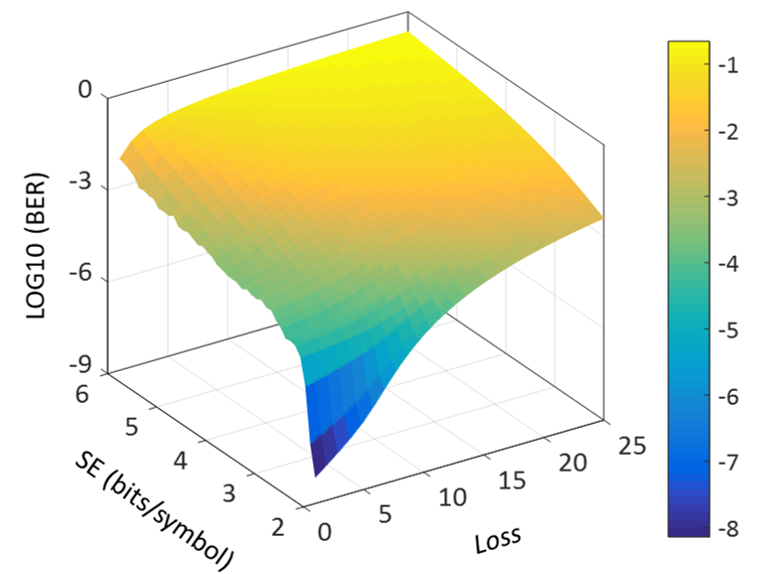}
\caption{Three-dimensional look-up table for BER, spectral efficiency (SE), and link loss under the optimal clipping ratio.}
\label{BERvsSEvsESNR}
\end{figure}

For the PS-64QAM signal, the spectral efficiency corresponds to the specific probabilities $\boldsymbol{Pr}$ of the amplitudes. As shown in Eqs. (\ref{eq17})-(\ref{eq19}), the $E_{\text{b},~i}$ is related to the $\boldsymbol{Pr}$ (i.e., spectral efficiency) and the $d$ where $i$ is from 1 to 3. Meanwhile, as Eq. (\ref{eq20}) shows, the $d$ is relevant to the $Loss_i$ and the $\boldsymbol{Pr}$ (i.e., spectral efficiency) under a given clipping ratio. In conclusion, the theoretical BER of the PS-64QAM signal is determined by the $Loss_i$ and spectral efficiency. Fig. \ref{BERvsSEvsESNR} depicts the three-dimensional look-up table for BER, spectral efficiency, and link loss under the optimal clipping ratio for the flexible PtMP optical networks. In Fig. \ref{CapacityvsCR}, the optimal clipping ratio can be confirmed depending on the link losses of the leaves to achieve the highest capacity limitation of the flexible PtMP optical networks. The three-dimensional look-up table is established by Eq. (\ref{eq21}) under the optimal clipping ratio. As Fig. \ref{BERperSubcarrier} shows, the BER of each subcarrier for the DSCM-based flexible PtMP optical networks with the clipping operation has a large margin compared to the 7\% FEC limit. Therefore, when the targeted BER is set to 7\% FEC limit, the spectral efficiency can be further improved for each subcarrier based on the three-dimensional look-up table.

\begin{algorithm}[!t]
\caption{Optimization strategy for the capacity of flexible point-to-multi-point optical networks}
\label{algorithm1}
\begin{algorithmic}[1]
\vspace{0.1cm}
\Require Subcarrier number $N$; link losses $\boldsymbol{Loss}$; Noise variance $\sigma^2_\text{n}$; Signal power $P_{\text{DSCM}}$; Targeted BER $BER_\text{T}$; Maximum spectral efficiency $SE_{\text{max}}$; Spectral-efficiency updating step size $\Delta SE$; Three-dimensional look-up tables $\boldsymbol{LUT}(BER,~SE,~Loss)$.
\vspace{0.1cm}
\Ensure Optimal clipping ratio $\eta_{\text{opt}}$ and optimal spectral efficiencies $\boldsymbol{SE}$.
\vspace{0.1cm}
\State Calculate the effective SNRs $\boldsymbol{ESNR}$ of the subcarriers by Eq. (\ref{eq10}) based on subcarrier number $N$, link losses $\boldsymbol{Loss}$, noise variance $\sigma^2_\text{n}$, and signal power $P_{\text{DSCM}}$.
\State Calculate the optimal clipping ratio $\eta_{\text{opt}}$ by Eq. (\ref{eq12}) based on the $\boldsymbol{ESNR}$.
\State Choose the corresponding three-dimensional look-up table $LUT (BER, SE, Loss)$ based on the optimal clipping ratio $\eta_{\text{opt}}$.
\Comment{Next, the iterative algorithm is used to find the optimal $\boldsymbol{SE}$ based on $BER_\text{T}$ and $\boldsymbol{Loss}$ in $LUT$.}
\State \textbf{Initialize:} $\boldsymbol{SE}:= SE_{\text{max}} \times \text{ones}(1, N)$
\For{$i = 1$ to $N$}
\State \textbf{Initialize:} $BER_i$ by $(SE_{i}, Loss_i)$ in $LUT$
\While{$BER_i > BER_\text{T}$}
\State Update $SE_i \gets SE_i - \Delta SE$
\State Update $BER_i$ by $(SE_i, Loss_i)$ in the $LUT$
\EndWhile
\EndFor
\State \textbf{return} Optimal clipping ratio $\eta_{\text{opt}}$ and optimal spectral efficiencies $\boldsymbol{SE}$
\end{algorithmic}
\end{algorithm}

\textbf{Algorithm \ref{algorithm1}} shows the proposed optimization strategy for increasing the capacity of the flexible PtMP optical networks. The input of the algorithm includes the subcarrier number $N$, link losses $\boldsymbol{Loss}$, noise variance $\sigma^2_\text{n}$, signal power $P_{\text{DSCM}}$, targeted BER $BER_\text{T}$, maximum spectral efficiency $SE_{\text{max}}$, spectral-efficiency updating step size $\Delta SE$, and three-dimensional look-up tables $\boldsymbol{LUT}(BER,~SE,~Loss)$. The first step of the proposed algorithm is to calculate the effective SNRs $\boldsymbol{ESNR}$ of the subcarriers by Eq. (\ref{eq10}) based on subcarrier number $N$, link losses $\boldsymbol{Loss}$, noise variance $\sigma^2_\text{n}$, and signal power $P_{\text{DSCM}}$. In the second step, the optimal clipping ratio $\eta_{\text{opt}}$ is calculated by Eq. (\ref{eq12}) based on the $\boldsymbol{ESNR}$. In the third step, the corresponding three-dimensional look-up table $LUT (BER, SE, Loss)$ is chosen based on the optimal clipping ratio $\eta_{\text{opt}}$. The last step uses an iterative algorithm to find the optimal $\boldsymbol{SE}$ based on $BER_\text{T}$ and $\boldsymbol{Loss}$ based on the $LUT$. In the iterative algorithm, the spectral efficiency of all subcarriers is initialized as the maximum spectral efficiency $SE_{\text{max}}$ (i.e., 6bit/symbol for 64QAM). The outer iterations traverse all the subcarriers to find the optimized spectral efficiencies $\boldsymbol{SE}$. In the inner iterations, the spectral efficiency $SE_i$ is reduced by the step size $\Delta SE$ each iteration until the BER updated by $(SE_i, Loss_i)$ using the $LUT$ is lower than the targeted BER $BER_\text{T}$. After outer and inner iterations, the optimal spectral efficiencies $\boldsymbol{SE}$ are obtained for all the subcarriers of the flexible PtMP optical network. Finally, the optimal clipping ratio $\eta_{\text{opt}}$ and the optimal spectral efficiencies $\boldsymbol{SE}$ are returned.

\begin{figure}[!t]
\centering
\includegraphics[width=\linewidth]{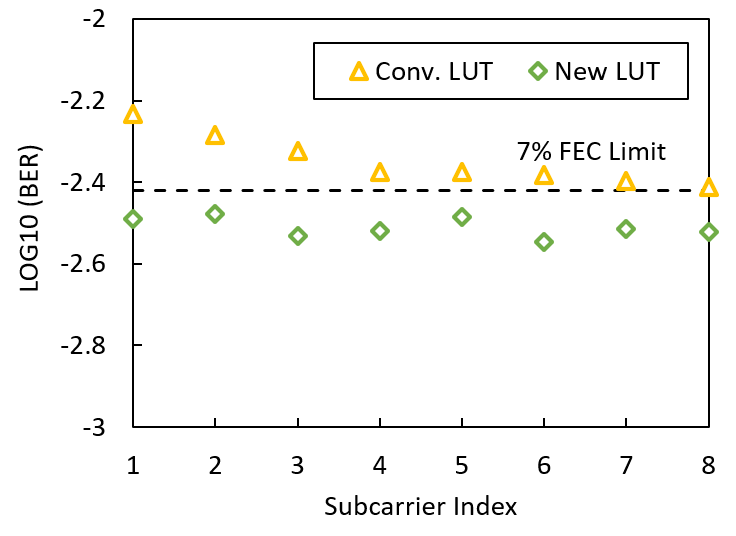}
\caption{The BERs of the subcarriers for the DSCM-based flexible PtMP optical networks after improving spectral efficiency by using a conventional look-up table (LUT) and a new look-up table.}
\label{BERperSubcarrierF}
\end{figure}

Figure \ref{BERperSubcarrierF} depicts the BERs of the subcarriers for the DSCM-based flexible PtMP optical networks after improving spectral efficiency by using a conventional look-up table and a new look-up table. For the conventional look-up table, the clipping noise is assumed to coincide with Gaussian distribution. Based on the conventional look-up table, the spectral efficiencies for the 8 subcarriers are set to [5.90, 5.72, 5.49, 5.19, 4.87, 4.54, 4.20, 3.90] bits/symbol, respectively. However, the BERs of all the subcarriers cannot be under the targeted 7\% FEC limit. Thus, the assumption of Gaussian distribution is not exact for the clipping noise. For the new look-up table, the clipping noise coincides with the distribution based on Eq. (\ref{eq13}). When the new look-up table in Fig. \ref{BERvsSEvsESNR} is used, the BERs of all the subcarriers are below and close to the targeted 7\% FEC limit. Therefore, the new look-up table is more precise than the conventional look-up table. 

When the clipping operation is not used, the spectral efficiencies for the 8 subcarriers are set to [4.80, 4.40, 4.00, 3.60, 3.20, 2.80, 2.40, 2.00] bits/symbol, respectively. The corresponding capacity is approximately 217.6Gb/s (i.e. 203.3GB/s excluding the 7\% FEC overhead) for the flexible PtMP optical networks based on 8$\times$8Gbaud DSCM signal. When the new look-up table is employed, the spectral efficiencies for the 8 subcarriers are set to [5.65, 5.51, 5.27, 5.04, 4.75, 4.38, 4.08, 3.80] bits/symbol, respectively. The corresponding capacity is approximately 307.84Gb/s (i.e. 287.7GB/s excluding the 7\% FEC overhead). Using the optimal clipping operation and new look-up table, the capacity of the DSCM-based flexible PtMP optical networks is approximately 90.4Gb/s higher than that without clipping operation, which approaches the theoretical result in Section \ref{SectionII}. Thus, the optimal clipping operation and new look-up table can be used to achieve 41\% higher capacity.

\section{Conclusion} \label{SectionIV}
In this paper, we derive the theoretical capacity limitation versus the link losses for the flexible PtMP optical networks based on the entropy-loading DSCM signal. Meanwhile, the clipping operation with the optimal clipping ratio reduces the PAPR of the entropy-loading DSCM signal to improve the effective SNR for approaching the highest capacity limit. Based on the piecewise power exponential model for the clipping noise, we establish a three-dimensional look-up table for bit-error ratio, spectral efficiency, and link loss under the optimal clipping ratio. Based on the three-dimensional look-up table, an optimization strategy is proposed to acquire optimal spectral efficiencies for achieving a higher capacity of the flexible PtMP optical networks.

\ifCLASSOPTIONcaptionsoff
  \newpage
\fi

\bibliographystyle{IEEEtran}
\bibliography{bibtex/bib/IEEEexample}

\end{document}